\documentstyle[preprint,aps]{revtex}

\title{Kaon production cross sections from baryon-baryon interactions}
\author{G. Q. Li and C. M. Ko}
\address{\it{
Cyclotron Institute and Physics Department\\  Texas A\&M University,
College Station, Texas 77843}}
\begin{document}
\maketitle

\begin{abstract}
In a one-pion plus one-kaon exchange model, we calculate the kaon production
cross sections in nucleon-nucleon, nucleon-delta and delta-delta interactions.
We find that this model describes reasonably well the experimental data on
kaon production in the proton-proton
interaction.  Near the kaon production threshold, the
cross section obtained from this model
is smaller than that from the linear parameterization
of Randrup and Ko. For
kaon production cross sections from the nucleon-delta and delta-delta
interactions, the cross sections are singular in free space, so we
calculate them in a nuclear medium by including the (complex)
pion self-energy. The results are compared with the scaling
ansatz of Randrup and Ko. The theoretical
cross sections are then used in a transport model to study
kaon production from Au+Au collisions at 1 GeV/nucleon.
\end{abstract}

\newpage

\section{introduction}

Kaon production in heavy-ion collisions has
been extensively studied both experimentally
\cite{EXP1,EXP2,EXP3,EXP4} and theoretically
\cite{RAN80,CUG84,AICH85,AICH87,ZWER87,MOS91,LI92,HUANG93,FANG94,MOS94,LIBA94,AI
CH94,LI95,ko91,ARC,RQMD,gqli95}
at various incident energies.
Its interest stems from the fact that because of strangeness
conservation, kaons will not be absorbed in nuclear medium
once they are produced in the collisions. They are thus expected to carry
useful information about the early stage of heavy-ion
collisions.  Indeed, at incident energies
below the kaon production threshold in the nucleon-nucleon interaction in
free space ($\approx$ 1.58 GeV), the kaon yield has been shown
in transport models
to be sensitive to both the nuclear equation of state at high densities
\cite{AICH85,MOS91,LI92,LIBA94,AICH94,LI95}
and the kaon scalar potential in nuclear medium \cite{FANG94}.
The latter is related to the partial restoration of chiral symmetry
\cite{kapl86,brow94}.
{}From comparing theoretical results with experimental data measured in
Au+Au collisions at 1 GeV/nucleon, it has been concluded that the
attractive kaon scalar potential in dense nuclear matter is appreciable
and the nuclear equation of state at high densities is soft.

For heavy ion collisions at incident energies of around 1 GeV/nucleon,
the colliding system consists mainly of nucleons, deltas and pions.
Kaons can be produced from baryon-baryon (nucleon-nucleon, nucleon-delta,
and delta-delta) and pion-baryon (pion-nucleon and pion-delta) interactions.
Kaon production cross sections in these elementary
processes are thus needed in transport models
in order to evaluate the kaon yield in heavy-ion collisions.
The most straightforward way is to use
experimentally measured cross sections.
There exist some experimental data for kaon production
cross sections \cite{DATA} from the pion-proton and proton-proton interactions.
For the proton-proton interaction, there are no data near the
the kaon production threshold of about 1.58 GeV, which are the energies
at which kaons are mostly produced in heavy ion collisions at around
1 GeV/nucleon.  The extrapolation from
the available experimental data at high energies
to near the threshold thus depends on the parameterization used
for the cross sections.
There are two popular parameterizations of the experimental data;
the linear parameterization of Randrup and Ko \cite{RAN80}
and the quartic parameterization of Zwermann \cite{ZWER87}.
Although both fit the experimental data at the lowest
available incident energy,
they differ considerably near the kaon production threshold.

Furthermore, the kaon production
cross sections in nucleon-delta and delta-delta interactions are
not available experimentally. In the work of Randrup and Ko \cite{RAN80},
these cross sections
are related to that from the nucleon-nucleon interaction mainly through
arguments based on isospin symmetry. This leads to a simple scaling
ansatz:
$\sigma _{NN\rightarrow NYK} (\sqrt s)=(4/3)\sigma _{N\Delta \rightarrow
NYK} (\sqrt s) =2 \sigma _{\Delta\Delta \rightarrow NYK} (\sqrt s)$.
No detailed theoretical study has been carried out to examine its
accuracy. All existing calculations based on transport models
have shown that deltas play a significant role in
subthreshold kaon production due to their larger masses than the nucleon mass
\cite{AICH85,MOS91,LI92,FANG94}.
In these calculations, the scaling ansatz
of Ref. \cite{RAN80} have usually been used.  At higher incident energies
in AGS experiments, deltas have also been found to
be important for kaon production. In microscopic
models such as the ARC \cite{ARC} and the RQMD \cite{RQMD},
the kaon production
cross sections from nucleon-delta and delta-delta interactions are
assumed, however,
to be the same as that from the nucleon-nucleon interaction.

The above discussions thus point to the need for a theoretical model
in evaluating the elementary kaon production cross sections in
hadron-hadron interactions.  In Ref. \cite{FAE94},
a resonance model has been used
to study kaon production from the pion-nucleon interaction
and has been shown to reproduce
rather well the available experimental data.
The resonance model has also been extended to kaon production
from the pion-delta interaction.
For kaon production in the nucleon-nucleon interaction,
a one-pion exchange model has been used
in Ref. \cite{KO89} by including also medium
modifications of the pion propagation in the delta-hole model.
The contribution of one-kaon exchange to kaon production in proton-proton
interaction has been studied in Ref. \cite{LAGET91}, and
has been found to be as important
as the one-pion exchange contribution if off-shell effects are
included. No
detailed calculation has been carried out
for kaon production from the nucleon-delta
and delta-delta interactions.

In this paper, we shall calculate kaon production cross sections in
nucleon-nucleon, nucleon-delta and delta-delta interactions based on
a one-pion plus one-kaon exchange model. In Section 2, we present the
formalism for these cross sections.
The results are presented in Sections 3 and compared with available
experimental data and existing parameterizations. In particular, we discuss
kaon production cross sections from the nucleon-delta and
delta-delta interactions and compare them with the scaling ansatz of Ref.
\cite{RAN80}.
In section 4, we use the theoretical elementary kaon production cross sections
in a transport model to study kaon
production from Au+Au collisions at 1 GeV/nucleon. Finally, a summary is
given in Section 5.

\section{Kaon production in baryon-baryon interactions: Formalism}

In this section, we present the formalism for
kaon production cross sections
in nucleon-nucleon, nucleon-delta and delta-delta interactions.
Only final states without a delta will be considered as the threshold
energies for final states with a delta
are much higher than the average energy of two interacting
baryons in heavy ion collisions at 1 GeV/nucleon.

\subsection{$NN\rightarrow NYK$}

The relevant pion and kaon exchange
diagrams are shown in Fig. 1, where
$Y$, denoted by thick solid lines, represents either a $\Lambda$ or
a $\Sigma$ hyperon. Let us first consider the case in which the kaon is
produced in association with a $\Lambda$ hyperon. To fix the relative sign
between the pion and kaon exchange amplitudes requires a well defined
interacting Lagrangian. Since we shall treat the $\pi NYK$ and $KNNK$
interactions phenomenologically using their cross sections determined
from the resonance model \cite{FAE94}, their relative sign can not be
fixed. Instead
choosing the sign to maximize the cross sections
as in Ref. \cite{LAGET91}, we shall simply neglect
the interference between the two amplitudes and fit the experimental
cross sections by adjusting other parameters in the model.
The isospin-averaged cross section is then given by \cite{KO89,YAO}

\begin{eqnarray}
&&\sigma _{NN\rightarrow N\Lambda K}(\sqrt s)=\nonumber\\
&&{3m_N^2\over 2\pi ^2p^2s}\int ^{\big(w_\pi\big)_{max}}_{\big(w_\pi\big)
_{min}} dw_\pi w_\pi^2 k_\pi
\int ^{(q^2_\pi)_+}_{(q^2_\pi)_-} dq_\pi^2 {f_{\pi NN}^2\over m_\pi ^2}
F^4(q^2_\pi) q^2_\pi \vert D_\pi(q_\pi^2)\vert^2 {\bar \sigma}_{\pi
N\rightarrow
\Lambda K}(w_\pi)\nonumber\\
&&+{m_Nm_\Lambda \over 2\pi ^2p^2s}\int ^{\big(w_K\big)_{max}}_{\big(w_K\big)
_{min}} dw_K w_K^2 k_K
\int ^{(q^2_K)_+}_{(q^2_K)_-} dq_K^2 {f_{KN\Lambda}^2\over m_K ^2}
F^4(q^2_K) q^2_K  \vert D_K(q^2_K)\vert^2 {\bar \sigma} _{KN\rightarrow KN}
(w_K).
\end{eqnarray}
In the above, masses of the nucleon, pion, kaon, and $\Lambda$ hyperon
are denoted
by $m_N,~m_\pi, m_K,$  and $m_\Lambda$, respectively. $f_{\pi NN}$ and
$f_{KN\Lambda}$ are the pseudovector $\pi NN$ and $KN\Lambda$
coupling constants.
The four-momenta of two initial nucleons are denoted by
$p_1$ and $p_2$, respectively. Their total energy in the center-of-mass
system is denoted by $\sqrt s$,
while the magnitude of the three-momentum of each nucleon
in the center-of-mass frame is $p$.
$w_\pi$ denotes the total energy
of the pion-nucleon system in the center-of-mass frame, with
$(w_\pi )_{min}=m_K+m_\Lambda$ and $(w_\pi )_{max}=\sqrt s-m_N$.
Similarly, $w_K$ is the
total energy of the kaon-nucleon system, with $(w_K)_{min}=m_K+m_N$ and
$(w_K)_{max}=\sqrt s-m_\Lambda $. The
magnitude of the three-momentum of the exchanged pion, $k_\pi$, is related
to $w_\pi$ by
\begin{eqnarray}
k_\pi ={1\over 2w_\pi}\sqrt{[w^2_\pi -(m_N+m_\pi)^2]
[w^2_\pi -(m_N-m_\pi )^2]},
\end{eqnarray}
and a similar expression for $k_K$, with $m_\pi$ and $w_\pi$
in Eq. (2) replaced by $m_K$ and $w_K$, respectively.

In Eq. (1), the four-momentum of the exchanged pion is denoted by $q_\pi$.
In the center-of-mass frame, we have
\begin{eqnarray}
(q_\pi^2)_- =2m_N^2-2(m_N^2+p^2)^{1/2}(m_N^2+p^{\prime 2})^{1/2}+2pp^\prime,
\nonumber\\
(q_\pi^2)_+ =2m_N^2-2(m_N^2+p^2)^{1/2}(m_N^2+p^{\prime 2})^{1/2}-2pp^\prime,
\end{eqnarray}
where $p^\prime$ is the magnitude of the three-momentum of the final
nucleon and is related to $\sqrt s$ and $w_\pi$ by
\begin{eqnarray}
p^\prime ={1\over 2}\sqrt{[s-(w_\pi +m_N)^2][s-(w_\pi-m_N)^2]/s}.
\end{eqnarray}
Similarly, for $(q^2_K)_{\pm}$ we have
\begin{eqnarray}
(q_K^2)_- =m_N^2+m_\Lambda ^2-2(m_N^2+p^2)^{1/2}(m_\Lambda^2
+p^{\prime 2})^{1/2}+2pp^\prime,
\nonumber\\
(q_K^2)_+ =m_N^2+m_\Lambda ^2-2(m_N^2+p^2)^{1/2}(m_\Lambda^2
+p^{\prime 2})^{1/2}-2pp^\prime,
\end{eqnarray}
with $p^\prime$ calculated from Eq. (4) by replacing $m_N$ with
$m_\Lambda$.

The pion propagator $D_\pi(q_\pi^2)$ and the kaon propagator $D_K(q_K^2)$
in free space are given, respectively, by
$$D_\pi(q_\pi^2)={1\over q_\pi^2-m_\pi^2},$$
and
$$D_K(q_K^2)={1\over q_K^2-m_K^2}.$$

In Eq. (1), ${\bar \sigma} _{\pi N \rightarrow\Lambda K}$ should be the
isospin-averaged $\pi N\rightarrow \Lambda K$
cross section for an off-shell pion,
but we replace it by an on-shell one. It can then be
obtained from the experimental data via
$${\bar \sigma} _{\pi N\rightarrow \Lambda K}
={1\over 2}\sigma _{\pi ^-p\rightarrow \Lambda K^0}.$$
It has been parameterized by Cugnon {\it et
al.} \cite{CUG84} and used in Ref. \cite{KO89}. It can also be calculated
from theoretical models.
We use here the results of Ref. \cite{FAE94} based
on a resonance model which describes rather well the experimental data.
To take into account the off-shell nature of the exchanged pion,
we introduce as in Ref. \cite{LAGET91}
at the $\pi N \Lambda K$ vertex a pion form factor
similar to the one used in the $\pi NN$ vertex, i.e.,
\begin{eqnarray}
F (q_\pi ^2) ={\Lambda _\pi ^2-m_\pi ^2\over
\Lambda _\pi^2-q_\pi ^2},\nonumber
\end{eqnarray}
with $\Lambda _\pi$ being the cut-off parameter.

In the kaon exchange contribution, ${\bar \sigma}
_{KN\rightarrow KN}$
is the isospin-averaged kaon-nucleon scattering cross section, and can be
determined from experiments by
$${\bar \sigma} _{KN\rightarrow KN} = {1\over 2}\big(\sigma _{K^+p\rightarrow
K^+p}+\sigma _{K^0p\rightarrow K^0p} +\sigma _{K^0p\rightarrow K^+n}\big),$$
where we have made use of the following relations based on the
isospin symmetry:
$\sigma _{K^0n\rightarrow K^0n}=\sigma _{K^+p\rightarrow K^+p},
{}~\sigma _{K^+n\rightarrow K^+n}=\sigma _{K^0p\rightarrow K^0p},$
and $\sigma _{K^+n\rightarrow K^0p}=\sigma _{K^0p\rightarrow K^+n}$.
These cross sections are taken from
the phase shift analysis of Martin \cite{MARTIN76}.
The off-shell effect is again included by the following
kaon form factor at both vertices,
\begin{eqnarray}
F (q_K ^2) ={\Lambda _K ^2-m_K ^2\over
\Lambda _K^2-q_K ^2},\nonumber
\end{eqnarray}
with $\Lambda _K$ being the cut-off parameter.

For the reaction $pp\rightarrow p\Lambda K^+$, where experimental data exist,
the cross section is similar to Eq. (1) except that the first term on the
right hand side is reduced by a factor of 3 due to the isospin and
$\bar\sigma_{KN\to KN}$ in the second term replaced by
$\bar\sigma_{K^+p\to K^+p}$.
If only the pion exchange
is included, the isospin-averaged cross section is three times
the $pp\rightarrow p\Lambda K^+$ cross section as
in Refs. \cite{RAN80,KO89}. Since the isospin factor
for the kaon exchange is one, the isopsin-averaged cross section
including both pion and kaon exchanges
is thus less than three times the $pp\rightarrow p\Lambda K^+$ cross
section.

Next, let us consider the reaction where a kaon is produced in association
with a $\Sigma $ hyperon. Since the $\Sigma$ hyperon has three charge states,
there are more final states than in the case for the $\Lambda $ hyperon. The
isospin-averaged cross section is given by

\begin{eqnarray}
&&\sigma _{NN\rightarrow N\Sigma K}(\sqrt s)=\nonumber\\
&&{3m_N^2\over 2\pi ^2p^2s}\int ^{\big(w_\pi\big)_{max}}_{\big(w_\pi\big)
_{min}} dw_\pi w_\pi^2 k_\pi
\int ^{(q^2_\pi)_+}_{(q^2_\pi)_-} dq_\pi^2 {f_{\pi NN}^2\over m_\pi ^2}
F^4(q^2_\pi) q^2_\pi \vert D_\pi(q^2_\pi)\vert^2 {\bar \sigma}_{\pi
N\rightarrow
\Sigma K}(w_\pi)\nonumber\\
&&+{3m_Nm_\Sigma \over 2\pi ^2p^2s}\int ^{\big(w_K\big)_{max}}_{\big(w_K\big)
_{min}} dw_K w_K^2 k_K
\int ^{(q^2_K)_+}_{(q^2_K)_-} dq_K^2 {f_{KN\Sigma}^2\over m_K ^2}
F^4(q^2_K) q^2_K \vert D_K(q^2_K)\vert^2 {\bar \sigma} _{KN\rightarrow
KN}(w_K).
\end{eqnarray}
The notations are similar to those in Eq. (1). The mass of the
$\Sigma $ hyperon and the
$KN\Sigma$ coupling constant
are denoted by $m_\Sigma$ and $f_{KN\Sigma}$, respectively.
The isospin-averaged
cross section for the $\pi N\rightarrow \Sigma K$ reaction is denoted
by ${\bar \sigma} _{\pi N\rightarrow \Sigma K}$ and is calculated from
$${\bar \sigma} _{\pi N\rightarrow \Sigma K}={1\over 3}\big(\sigma _{\pi
^+p\rightarrow \Sigma ^+k^+} +2\sigma _{\pi ^+ n\rightarrow \Sigma ^0K^+}
+\sigma _{\pi ^0p\rightarrow \sigma ^0K^+} + \sigma _{\pi ^-p \rightarrow
\Sigma ^0K^+}).$$
For the cross sections on the right hand side,
we use again the results of Ref. \cite{FAE94} based
on the resonance model which has been shown to fit the experimental data
satisfactorily.

For $pp\rightarrow n\Sigma ^+K^+$, $pp\rightarrow
p\Sigma ^0 K^+$, and $pp\rightarrow p\Sigma ^+K^0$,
there are experimental data available, and their cross sections can be
obtained from Eq. (6) by replacing
$3\bar\sigma_{\pi N\to\Sigma K}$ in the first term by
$2\bar\sigma_{\pi^+p\to\Sigma^+K^+}$, $\bar\sigma_{\pi^0p\to\Sigma^0K^+}$,
and $\bar\sigma_{\pi^0p\to\Sigma^+K^0}$, respectively, and
$3\bar\sigma_{KN\to KN}$ in the second term
by $2\bar\sigma_{K^0p\to K^+n}$, $\bar\sigma_{K^+p\to K^+p}$,
and $2\bar\sigma_{K^0p\to K^0p}$, respectively.

\subsection{$N\Delta\rightarrow NYK$}

As mentioned in the Introduction, deltas play an important role in
subthreshold kaon production from heavy-ion collisions. Unfortunately,
there are no experimental data for kaon production
cross sections from the nucleon-delta and delta-delta interactions. A
theoretical model is therefore needed to determine these
cross sections. In this and next subsections we extend the pion
and kaon exchange model to calculate kaon production cross sections
in these reactions.

For the $N\Delta\rightarrow NYK$ reaction, the relevant diagrams
for pion and kaon exchanges are shown in
Fig. 2. In each case, there are two different diagrams; one with
a kaon produced from the meson-nucleon interaction and
the other from the meson-delta interaction.
For $N\Delta\rightarrow N\Lambda K$ reaction, diagram (d) does
not exist as there
is no $K\Delta \Lambda$ coupling due to isospin conservation.
The isospin-averaged cross section for this reaction is

\begin{eqnarray}
&&\sigma _{N\Delta\rightarrow N\Lambda K}(\sqrt s)=\nonumber\\
&&{3m_N^2\over 4\pi ^2p^2s}\int ^{\big(w_\pi\big)_{max}}_{\big(w_\pi\big)
_{min}} dw_\pi w_\pi^2 k^\prime_\pi
\int ^{(q^2_\pi)_+}_{(q^2_\pi)_-} dq_\pi^2 {f_{\pi NN}^2\over m_\pi ^2}
F^4(q_\pi^2)q^2_\pi \vert D_\pi(q_\pi ^2)\vert^2
{\bar \sigma}_{\pi \Delta\rightarrow \Lambda K}(w_\pi)\nonumber\\
&&+{m_Nm_\Delta\over 4\pi ^2p^2s}\int ^{\big(w_\pi\big)_{max}}_{\big(w_\pi\big)
_{min}} dw_\pi w_\pi^2 k_\pi
\int ^{(q^2_\pi)^\prime_+}_{(q^2_\pi)^\prime_-}
dq_\pi^2 {f_{\pi N\Delta}^2\over m_\pi ^2}
F^4(q^2_\pi) A(q_\pi ^2) \vert D_\pi(q_\pi^2)\vert^2
{\bar \sigma} _{\pi N\rightarrow\Lambda K}(w_\pi)\nonumber\\
&&+{m_Nm_\Lambda \over 4\pi ^2p^2s}\int ^{\big(w_K\big)_{max}}_{\big(w_K\big)
_{min}} dw_K w_K^2 k_K^\prime
\int ^{(q^2_K)_+}_{(q^2_K)_-} dq_K^2 {f_{KN\Lambda}^2\over m_K ^2}
F^4(q^2_K) q^2_K \vert D_K(q^2_K)\vert^2
{\bar \sigma} _{K\Delta\rightarrow KN}(w_K).
\end{eqnarray}
The notations are again
similar to those in Eq. (1). In the first term, corresponding
to Fig. 2(a), $k_\pi^\prime$ is calculated from Eq. (2) with $m_N$ replaced
by the mass of the delta resonance, $m_\Delta$. ${\bar \sigma} _{\pi \Delta
\rightarrow \Lambda K}$ is the isospin-averaged cross section for the
$\pi \Delta\rightarrow \Lambda K$ reaction which is determined from
$${\bar \sigma} _{\pi\Delta\rightarrow \Lambda K}={1\over 3}
\sigma _{\pi ^-\Delta ^{++}\rightarrow \Lambda K^+},$$
with $\sigma _{\pi ^-\Delta ^{++}\rightarrow \Lambda K^+}$
taken from Ref. \cite{FAE94}.

In the second term, corresponding to Fig. 2(b), $f_{\pi N\Delta}$
is the $\pi N\Delta$ coupling constant, and $(q_\pi ^2)^\prime _{\pm}$
are given by
\begin{eqnarray}
(q_\pi^2)^\prime_- =m_N^2+m_\Delta ^2
-2(m_\Delta^2+p^2)^{1/2}(m_N^2+p^{\prime 2})^{1/2}+2pp^\prime,
\nonumber\\
(q_\pi^2)^\prime_+ =m_N^2+m_\Delta ^2
-2(m_\Delta^2+p^2)^{1/2}(m_N^2+p^{\prime 2})^{1/2}-2pp^\prime.
\end{eqnarray}
$A(q_\pi^2)$ is from the $\pi N\Delta$ vertex and is given by
$$A(q_\pi ^2)={1\over 48 m_Nm_\Delta}\nonumber\\
\big[q_\pi ^2-(m_\Delta +m_N)^2\big]
\big[(m_\Delta ^2-m_N^2+q_\pi ^2)^2-4m_\Delta ^2q_\pi ^2\big].$$

In the third term, corresponding to Fig. 2(c), $k_K^\prime$ is calculated
from Eq. (2) with $m_N$ and $m_\pi$
replaced by $m_\Delta$ and $m_K$, respectively.
${\bar \sigma}_{K\Delta \rightarrow KN}$ is the isospin-averaged
cross section for the $K\Delta\rightarrow KN$ reaction.
Since there is neither experimental nor theoretical information on this
cross section, we assume that it is the same as
the isospin-averaged cross section
${\bar \sigma} _{KN\rightarrow KN}$ for the
$KN\rightarrow KN$ reaction.

Since a delta can decay into a physical pion and a nucleon,
the exchanged pion in Fig. 2(b) can be on shell, which
leads to a singularity in the pion propagator.  However,
a pion in nuclear matter
acquires a (complex) self-energy due to the strong pion-nucleon interaction.
The finite imaginary part of the pion self-energy then
makes the contribution from the on-shell pion finite.
Since we are interested in kaon production in nuclear medium, we
replace the free space pion propagator by an in-medium one,
\begin{eqnarray}
D(q_\pi^2)={1\over q_\pi^2-m_\pi^2-\Pi},
\end{eqnarray}
with $\Pi$ being the pion self-energy to be specified later.

The $\Delta N\to NYK$ reaction includes contributions from both
on-shell and off-shell pions.
In transport models, the on-shell pion contribution
is usually treated as a two step process, i.e.,
a delta decaying into a physical pion
and a nucleon, and the subsequent production of a kaon from
the pion-nucleon interaction \cite{XIONG90}.  In this case,
we should include only
the off-shell pion contribution in $\Delta N\to NYK$.
Since it is not clear how one can properly
separate the on-shell and the off-shell
contribution in $\Delta N\to NYK$, we shall use
in the transport model the total $\Delta N\to NYK$ cross section
and neglect kaon production from explicit
pion-nucleon interactions. A similar strategy has been adopted
in Ref. \cite{MOS95} where the eta $(\eta )$ production cross
section from the nucleon-delta interaction has been considered.

For $N\Delta\rightarrow N\Sigma K$, the isospin-averaged cross section
is given by
\begin{eqnarray}
&&\sigma _{N\Delta\rightarrow N\Sigma K}(\sqrt s)=\nonumber\\
&&{3m_N^2\over 4\pi ^2p^2s}\int ^{\big(w_\pi\big)_{max}}_{\big(w_\pi\big)
_{min}} dw_\pi w_\pi^2 k^\prime_\pi
\int ^{(q^2_\pi)_+}_{(q^2_\pi)_-} dq_\pi^2 {f_{\pi NN}^2\over m_\pi ^2}
F^4(q^2_\pi) q^2_\pi \vert D_\pi(q_\pi^2)\vert^2 {\bar \sigma}_{\pi
\Delta\rightarrow \Sigma K}(w_\pi)\nonumber\\
&&+{m_Nm_\Delta\over 4\pi ^2p^2s}\int ^{\big(w_\pi\big)_{max}}_{\big(w_\pi\big)
_{min}} dw_\pi w_\pi^2 k_\pi
\int ^{(q^2_\pi)^\prime_+}_{(q^2_\pi)^\prime_-}
dq_\pi^2 {f_{\pi N\Delta}^2\over m_\pi ^2}
F^4(q^2_\pi)A(q_\pi^2) \vert D_\pi(q_\pi^2)\vert^2{\bar \sigma} _{\pi
N\rightarrow
\Sigma K}(w_\pi)\nonumber\\
&&+{3m_Nm_\Sigma \over 4\pi ^2p^2s}\int ^{\big(w_K\big)_{max}}_{\big(w_K\big)
_{min}} dw_K w_K^2 k_K^\prime
\int ^{(q^2_K)_+}_{(q^2_K)_-} dq_K^2 {f_{KN\Lambda}^2\over m_K ^2}
F^4(q^2_K) q^2_K \vert D_K(q^2_K)\vert^2 {\bar \sigma} _{K\Delta\rightarrow
KN}(w_K)\nonumber\\
&&+{m_\Delta m_\Sigma \over 4\pi ^2p^2s}\int ^{\big(w_K\big)_{max}}_
{\big(w_K\big)_{min}} dw_K w_K^2 k_K
\int ^{(q^2_K)^\prime_+}_{(q^2_K)^\prime_-} dq_K^2 {f_{K\Delta\Sigma}^2
\over m_K ^2}F^4(q^2_K)
A(q_K^2) \vert D_K(q^2_K)\vert^2{\bar \sigma} _{KN\rightarrow KN}(w_K).
\end{eqnarray}
Similar notations as those in Eq. (11) are used. $A(q_K^2)$, obtained
from the $K\Delta\Sigma$ vertex, is given by
$$A(q_K^2)=
{1\over 48m_\Sigma m_\Delta ^3}\nonumber\\
\big[q_K^2-(m_\Delta +m_\Sigma )^2\big]
\big[(m_\Delta ^2-m_\Sigma ^2+q_K^2)^2-4m^2_\Delta q_K^2\big].$$
The isospin-averaged
cross section ${\bar \sigma} _{\pi\Delta\rightarrow \Sigma K}$
is determined from
$${\bar \sigma} _{\pi\Delta\rightarrow \Sigma K}= {1\over 6}
\big( \sigma _{\pi ^+\Delta ^0\rightarrow \Sigma ^0K^+}
+{4\over 3}\sigma _{\pi ^+\Delta ^-\rightarrow \Sigma ^-K^+}
+\sigma _{\pi ^0\Delta ^0\rightarrow \Sigma ^-K^+}
+{5\over 3}\sigma _{\pi ^-\Delta ^{++}\rightarrow \Sigma ^0K^+}\big).$$
The cross sections on the right hand side are again
taken from Ref. \cite{FAE94}.
For $N\Delta\to N\Sigma K$, a kaon can also be produced from the
the $KN$ vertex (Fig. 2(d)). This contribution is given by the last term
in Eq. (10), where $f_{K\Delta\Sigma}$ is the $K\Delta\Sigma$ coupling
constant, and $(q^2_K)^\prime _{\pm}$ are calculated from Eq. (5) with
$m_N$ and $m_\Lambda$ replaced by $m_\Delta$ and $m_\Sigma$, respectively.

\subsection{$\Delta\Delta\rightarrow NYK$}

The pion and kaon exchange diagrams for this process are shown
in Fig. 3. In the case a kaon is produced in association with a
$\Lambda$ hyperon, the kaon exchange is not allowed
because of isospin conservation. The isospin-averaged cross section
is given by

\begin{eqnarray}
&&\sigma _{\Delta\Delta\rightarrow N\Lambda K}(\sqrt s)=\nonumber\\
&&{m_N m_\Delta\over 2\pi ^2p^2s}\int ^{\big(w_\pi\big)_{max}}_{\big(w_\pi\big)
_{min}} dw_\pi w_\pi^2 k^\prime _\pi
\int ^{(q^2_\pi)^\prime_+}_{(q^2_\pi)^\prime_-}
dq_\pi^2 {f_{\pi N\Delta}^2\over m_\pi ^2}
F^4(q^2_\pi)A(q_\pi^2)\vert D_\pi(q_\pi^2)\vert^2
{\bar \sigma} _{\pi \Delta\rightarrow
\Lambda K}(w_\pi).
\end{eqnarray}

On the other hand, both pion and kaon exchanges contribute
to the $\Delta\Delta\rightarrow N\Sigma K$ reaction.
The isospin-averaged cross section is

\begin{eqnarray}
&&\sigma _{\Delta\Delta\rightarrow N\Sigma K}(\sqrt s)=\nonumber\\
&&{m_Nm_\Delta\over 2\pi ^2p^2s}\int ^{\big(w_\pi\big)_{max}}_{\big(w_\pi\big)
_{min}} dw_\pi w_\pi^2 k^\prime_\pi
\int ^{(q^2_\pi)^\prime_+}_{(q^2_\pi)^\prime_-}
dq_\pi^2 {f_{\pi N\Delta}^2\over m_\pi ^2}
F^4(q^2_\pi)A(q_\pi^2)\vert D_\pi(q_\pi^2)\vert^2
{\bar \sigma} _{\pi \Delta\rightarrow
\Sigma K}(w_\pi)\nonumber\\
&&+{m_\Delta m_\Sigma \over 2\pi ^2p^2s}\int ^{\big(w_K\big)_
{max}}_{\big(w_K\big)_{min}} dw_K w_K^2 k^\prime_K
\int ^{(q^2_K)^\prime_+}_{(q^2_K)^\prime_-} dq_K^2 {f_{K\Delta\Sigma}^2
\over m_K ^2}F^4(q^2_K)
A(q_K^2) \vert D_K(q^2_K)\vert^2{\bar \sigma} _{K\Delta\rightarrow KN}(w_K).
\end{eqnarray}

\subsection{The pion self-energy in nuclear matter}

As mentioned earlier, to calculate the pion exchange diagrams for kaon
production in nucleon-delta and delta-delta interactions
requires the inclusion of a
(complex) pion self-energy. Following Refs.
\cite{KO89,brow75,PAN81,BROWN88,XIA89},
we calculate the self-energy of a pion with energy $\omega$ and
momentum ${\bf k}$ in nuclear matter by taking into account
the delta-hole polarization in the random-phase approximation, i.e.,
\begin{eqnarray}
\Pi (\omega ,{\bf k}) = {{\bf k} ^2 \chi (\omega,
{\bf k} )\over 1-g^\prime \chi (\omega ,{\bf k})},
\end{eqnarray}
where $g^\prime \approx 0.6$ is the Migdal parameter due to
the short-range repulsion.  The pion susceptibility $\chi$ is given by
\begin{eqnarray}
\chi (\omega ,{\bf k})\approx{8\over 9}
\Big({f_{\pi N \Delta}\over m_\pi }\Big)^2
{\omega _\Delta\over \omega ^2 -\omega _\Delta ^2}
{\rm exp}\Big(-2{\bf k}^2/b^2\Big) \rho ,
\end{eqnarray}
where $\rho$ is the nuclear matter density and
$b\approx 7m_\pi$ is the width of the form factor
\cite{PAN81}. Including the delta width $\Gamma_\Delta$,
$\omega _\Delta$ is
approximately given by
$$\omega _\Delta \approx{{\bf k}^2\over 2m_\Delta} + m_\Delta -m_N-{i\over 2}
\Gamma _\Delta.$$
The delta width in nuclear matter can be determined
by extending the delta-hole model to include effects of
the in-medium pion dispersion relation, i.e.,
\begin{eqnarray}
\Gamma _\Delta (\omega )&= - 2\int {d{\bf k}^\prime \over (2\pi )^3}
\Big({f_{\pi N\Delta}\over m_\pi }\Big)^2{\rm exp}\big(-2{\bf k}^{\prime 2}
/b^2\big)\nonumber\\
&\cdot{\rm Im}\Big[{{\bf k}^{\prime 2}\over 3}{1\over 1-g^\prime \chi }
D_\pi(\omega ,
{\bf k}^\prime ) {1\over 1-g^\prime \chi } +g^{\prime 2}{\Pi \over
{\bf k}^{\prime 2}}\Big],
\end{eqnarray}
where $D_\pi(\omega ,{\bf k})$ is the in-medium pion propagator
given by Eq. (9).

The delta decay width and the pion self-energy are then
calculated self-consistently.
We show in Fig. 4 the real (left panel) and imaginary (right panel) parts of
the
pion self-energy as  functions of its momentum for a number of energies.
The nuclear matter density is taken to be 2$\rho _0$, with $\rho _0$ =
0.17 fm$^{-3}$. We see that with increasing pion energy, the real
part decreases, while the imaginary part first increases and then decreases.

The pion self-energy is also affected by temperature. For heavy-ion collisions
at 1 GeV/nucleon, the temperature reached in the collisions is below
100 MeV, and according to Ref. \cite{henn94} its effect
is not appreciable. Thus, we shall not consider the temperature
effect in present study.

\section{kaon production in baryon-baryon interactions: results and
discussions}

Our model for kaon production in baryon-baryon interactions involves five
coupling constants and two cut-off parameters. We use the following
values for these parameters,

$${f_{\pi NN}^2\over 4\pi }= 0.08, ~~{f_{\pi N\Delta}^2\over 4\pi}
= 0.37,  ~~{f_{KN\Lambda}^2\over 4\pi} =0.97, ~~{f_{KN\Sigma}^2\over 4\pi}
= 0.07, ~~ {f_{K\Delta\Sigma}^2\over 4\pi}= 0.23,$$
$$\Lambda _\pi =1.2 ~{\rm GeV},
{}~~\Lambda _K =0.9 ~{\rm GeV}. $$
The $\pi NN$ coupling constant and the pion cut-off mass are typical
values used in the Bonn model for the nucleon-nucleon interaction
\cite{MACH87}. The $\pi N\Delta$ coupling constant are taken
from Ref. \cite{WEISE} and is consistent with that determined from
the $\Delta $ decay width.
The $KN\Lambda$ and $KN\Sigma$ coupling
constants are the same as those used in Ref. \cite{LAGET91}, and are close
to the values used in the Nijmegen  \cite{SWART}
as well as the Bonn-J\"urich \cite{SPETH}
models for the hyperon-nucleon interaction.
The $K\Delta\Sigma$ coupling constant is taken from Ref. \cite{SPETH}.
Finally, the kaon cut-off mass is similar to that used in Ref. \cite{LAGET91}.
The two cut-off parameters have been slightly adjusted in order to obtain
good agreements with available experimental data on kaon production.

The comparison of our model prediction with the experimental data for
$pp\rightarrow p\Lambda K^+$ is shown in Fig. 5,
where the solid curve gives the  result
including both the pion and kaon exchange contributions,
while the dash-dotted curve gives only
the pion-exchange contribution. We find as in Ref. \cite{LAGET91}
that once the off-shell effect at the $\pi N\Lambda K$ vertex is included,
the pion exchange alone
underestimates appreciably the experimental data. Thus, the
kaon exchange mechanism is essential for a correct account of the data.
As already pointed out in Refs. \cite{FANG94,AICH94},
the fact that the pion exchange alone also fits the experimental
data in Ref. \cite{KO89} is attributed to the neglect of
off-shell effects at the $\pi N\Lambda K$ vertex.

In Fig. 5, we also show
the predictions from the linear parameterization of Ref. \cite{RAN80}
(dashed curve)
and the quartic parameterization of Ref. \cite{ZWER87} (dotted curve),
which are given, respectively, by
$$\sigma _{linear} = 24\,{p_{max}\over m_K}  ~\mu b,$$
$$\sigma _{quartic} = 800\, ({p_{max}\over m_K})^4  ~\mu b,$$
with
\begin{eqnarray}
p_{max}={1\over 2\sqrt s}\Big[\big(s-(m_N+m_\Lambda +m_K)^2\big)
\big(s-(m_N+m_\Lambda -m_K)^2\big)\Big]^{1/2}.
\end{eqnarray}
At low energies (below the lowest experimental point of $\sqrt s \approx
2.7$ GeV) our prediction is close to the quartic parameterization, and is
considerably smaller than the linear parameterization.
Therefore, the use of the linear parameterization for subthreshold
kaon production in heavy-ion collisions probably overestimates the
contribution from the
nucleon-nucleon interaction. On the other hand, the quartic
parameterization greatly overestimates the cross section for $\sqrt s\ge 2.7$
GeV.

Figs. 6-8 show comparisons of our results with the experimental data
from reactions in which the kaon is produced in association with a
$\Sigma $ hyperon. In this case, the pion exchange alone can
explain the data as the kaon exchange contribution is very small
due to a small $KN\Sigma$ coupling.
This is consistent with the observations in Ref. \cite{LAGET91}.

In transport models, isospin-averaged cross sections are usually used.
We denote, by $\sigma _{BB\rightarrow NYK}$, the isospin-averaged kaon
production cross section in baryon-baryon interactions including both
the $N\Lambda K$ and $N\Sigma K$ final states, i.e.,
$\sigma _{BB\rightarrow NYK}=
\sigma _{BB\rightarrow N\Lambda K}+\sigma _{BB\rightarrow N\Sigma K}$.
In Fig. 9, we compare
$\sigma _{NN\rightarrow NYK}$ based on the present model (Eqs. (1) and (6))
with the parameterization of Ref. \cite{RAN80} (dashed curve)
\begin{eqnarray}
\sigma _{RK} =72\,{p_{max}\over m_K}+ 72\,{p^\prime_{max}\over m_K}\,\mu b,
\end{eqnarray}
where $p^\prime_{max}$ is calculated from Eq. (16) with $m_\Lambda$ replaced
by $m_\Sigma$.

The theoretical results for $\sigma_{NN\to NYK}$
are shown for nuclear matter at densities $\rho=
0, ~\rho_0$ and $2\rho_0$. In general, our results are
smaller than the linear parameterization of Ref. \cite{RAN80}, especially
near the threshold. The sudden change of the kaon cross section
at $\sqrt s\sim 2.63$GeV
in the linear parameterization is due to the onset of the
$NN\to N\Sigma K$ channel.
With increasing density, the kaon production
cross section is seen to increase.   This is consistent with
the findings of Ref. \cite{KO89}, and is mainly due to the
softening of the pion dispersion relation, which
tends to enhance the cross section.

Our results for the kaon production cross section in
the $N\Delta$ interaction is shown in Fig. 10.
The result from the scaling ansatz of Ref. \cite{RAN80}
(i.e., three-fourth of Eq. (17))
is also shown in the figure by the dashed curve. Our
results near the threshold are close to the scaling ansatz.
The density dependence of $\sigma_{N\Delta\to NYK}$
is, however, different from that of $\sigma_{NN\to NYK}$. While
the latter increases with density because of the softening of the
pion dispersion relation, the former decreases with density.
This is largely due to the strong suppression of the
singular on-shell pion contribution at high densities.
A similar observation has been found in Ref.
\cite{MOS95} where the eta production cross section in
the nucleon-delta interaction
is seen to also decrease with increasing density.

Finally, our results for the kaon production cross section in the
delta-delta interaction are shown in Fig. 11.
Near the threshold they are smaller
than the scaling ansatz of Ref. \cite{RAN80}
(i.e., half of Eq. (17)). Also, the kaon
production cross section in the delta-delta interaction is considerably smaller
than that in the nucleon-delta interaction as a result of the smaller
$\pi \Delta\rightarrow YK$ cross section than the
$\pi N\rightarrow YK$ cross section \cite{FAE94}.
Again, we see that the kaon production cross section decreases with
increasing density.

\section{Applications to Au+Au collisions at 1 GeV/nucleon}

In this section, we shall use the theoretical
kaon production cross sections from
baryon-baryon interactions in a transport model to study
kaon production in Au+Au collisions at 1 GeV/nucleon. This reaction has been
previously studied by many groups
\cite{HUANG93,FANG94,MOS94,LIBA94,AICH94,LI95}
based on various transport models and using mainly the
linear parameterization of Ref. \cite{RAN80} for the elementary kaon
production cross sections in baryon-baryon interactions.

Ideally, we would like to carry out the calculation in the
relativistic transport model as in our previous studies \cite{FANG94,LI95}.
In this way, we can treat properly the medium modification of hadron
properties \cite{FANG94}. Since the kaon production
cross sections from baryon-baryon interactions
determined in the present work have been obtained
without these medium effects, we shall use instead the non-relativistic
transport model in which hadron masses are taken to be free ones.
The difference between the relativistic and the
non-relativistic transport model for kaon production is, however,
insignificant, as already discussed
in Ref. \cite{LI95}. The reason is simple: in the reaction
$BB\rightarrow NYK$, both scalar and vector potentials in
the intial and final states cancel in leading order.
The kaon production threshold in the medium is thus not affected by medium
effects, which are neglected in the non-relativistic transport model.
The situation is certainly different for
$BB\rightarrow NNK{\bar K}$ and $BB\rightarrow NNp{\bar p}$
where a net scalar potential is present in the final state.
Including the change of the production threshold in the medium,
it has been shown that the relativistic transport model
leads to almost two order-of-magnitude enhancement of
the antiproton yield \cite{LI94B} and a factor of
4 enhancement of the antikaon yield
\cite{LI94C} as compared to those from the non-relativistic
transport model.  In the future, we plan to carry out
a more complete study of subthreshold kaon production in heavy ion
collisions based on the relativistic transport model
that includes consistently medium effects on the elementary kaon
production cross sections and the collision dynamics.

With a soft Skyrme equation of state, corresponding to
a compressibility $K=200$ MeV at normal nuclear matter density,
we show in Fig. 12 the energy distribution of baryon-baryon
interactions that lead to kaon production in
central Au+Au collisions at 1 GeV/nucleon. There are about 112
baryon-baryon collisions with available
energies above the respective kaon production thresholds. The average
energy in these collisions is about 2.64 GeV.
Since the threshold energy for
the $BB\rightarrow \Delta YK$ reaction with a delta in the final state
is about 2.84 GeV, our neglect of its contribution
to the kaon yield is therefore justified.
We would like to point out that
an average energy of 2.64 GeV corresponds to a kaon
maximum momentum of 0.275 GeV, which is very close to the average
kaon maximum momentum of 0.272 GeV obtained in Ref. \cite{LI95}
based on the relativistic transport model and a soft (relativistic)
equation of state. This is an indication of the similarity between the
relativistic and non-relativistic transport model descriptions of kaon
production in heavy-ion collisions.

The kaon production probabilities from different reactions
are shown in Fig. 13.
The results based on the theoretical elementary cross sections are
shown in the left panel.
The contribution from the nucleon-delta interaction is
most important and accounts for about 65\%. As in previous studies,
deltas play the most important role as
reactions involving delta resonances produce about 90\% of the kaons.

We have also carried out a VUU
calculation using the Randrup-Ko parameterization for kaon production
in baryon-baryon interactions. The results are shown in the middle panel
of Fig. 13. We see that the kaon yield in this case is only about 20\%
less than that obtained with the theoretical cross sections calculated
in the present paper.

To see quantitatively the difference (or similarity)
between the relativistic and
non-relativistic descriptions of kaon production
in heavy-ion collisions, we reproduce the Fig. 8 of Ref. \cite{FANG94}
in the right panel of Fig. 12. The relativistic transport
model is seen to lead to a reduction of about 30\% in the total kaon yield.
This reduction is mainly due to the momentum-dependent
nuclear mean-field potential included in the
relativistic transport model. This momentum-dependent potential gives
rise to an additional repulsion besides that from the compressional pressure,
and a smaller maximum density is thus reached in the relativistic
transport model. This reduction in kaon yield
is much smaller than that found in Ref. \cite{AICH87} based on
the non-relativistic QMD model with a momentum-dependent Skyrme-type
potential, where the reduction factor is about 3.
The reason for this difference between these
two calculations has been discussed in Ref. \cite{LI95}.
We believe that the treatment of the momentum-dependent mean-field potential
in Ref. \cite{AICH87} is incomplete as it has not taken into account
the difference in the intial and final potential
energies in the $BB\rightarrow NYK$ reaction.
Since baryons in the initial state have larger momenta,
the intial potential energy in the above reaction is larger
than that of the final state and can thus be used for kaon production.
If this potential difference is properly included, as in the relativistic
transport model, the difference between the kaon yields
obtained in QMD calculations with and without
the momentum-dependent potential should be small.

Finally, we show in Fig. 14 the kaon momentum
spectrum obtained in this study.
The theoretical result is seen to overestimate the experimental
data \cite{EXP3} by about 30\%. As mentioned earlier, the relativistic
transport model would lead to a reduction of a similar magnitude in
the kaon yield as compared to the non-relativistic one.
We thus expect that a consistent
calculation based on the relativistic transport model and using the
theoretical elementary cross sections which properly include the medium
effects, will give a better account of the experimental data.

\section{summary}

In this work, we have extended our previous studies of kaon production
in Au+Au collisions at 1 GeV/nucleon by using in the transport model
the elementary kaon production cross sections calculated from a theoretical
model. The main purpose of this
study is to examine our previous conclusions concerning
the nuclear equation of state at high densities \cite{LI95}
and the kaon scalar potential \cite{FANG94} in dense matter.
A particular concern is the possible underestimate of the
kaon production cross section from the nucleon-delta interaction
in the scaling ansatz of Randrup and Ko.
For this purpose, we have constructed
a one-pion plus one-kaon exchange model for kaon production
in baryon-baryon interactions. The parameters
of the model are fitted to available experimental data in
the proton-proton interaction. This model is
then extended to the nucleon-delta and delta-delta interactions.
We have found that near the kaon production threshold,
which are relevant for subthreshold kaon production, the cross sections
in the Randrup-Ko parameterization and scaling ansatz
that have been used in previous studies by us \cite{FANG94,LI95}
and by other groups \cite{HUANG93,LIBA94,AICH94} are overestimated
for the nucleon-nucleon interaction but underestimated for the nucleon-delta
interaction.

We have then used these theoretical cross sections
to calculate in a non-relativistic transport model
the kaon yield in Au+Au collisions at 1 GeV/nucleon.
The results are found in reasonable agreement with
experimental data. We expect that a consistent relativistic transport model
calculation will lead to an even better agreement with the data. The conclusion
of this extensive study is thus clear: we need both an attractive scalar
potential and a soft nuclear equation of state to account
for the experimental data from the Kaos collaboration at the GSI.

\medskip
We are grateful to J. Aichelin, G. E. Brown, B. A. Li, and K. Tsushima
for helpful discussions.
This work was supported in part by NSF Grant No. PHY-9212209
and the Welch Foundation Grant No. A-1110.

\newpage
\medskip\medskip
\centerline{\bf Figure Captions}

\begin{description}
\item{Fig. 1:} One-pion and one-kaon exchange diagrams for the $NN\rightarrow
NYK$ reaction.

\item{Fig. 2:} Same as Fig. 1, for the $N\Delta\rightarrow NYK$ reaction.

\item{Fig. 3:} Same as Fig. 1, for the $\Delta\Delta\rightarrow NYK$ reaction.

\item{ Fig. 4:} Real (left panel) and imaginary (right panel) parts of the
pion self-energy in nuclear matter with density $\rho =2 \rho _0$.

\item{Fig. 5:} Comparisons of model predictions with experimental data
for the kaon production cross section in $pp\rightarrow p\Lambda K^+$.
The dashed-dotted curve gives the one-pion exchange contribution, while
the solid curve includes both the one-pion and one-kaon exchange
contributions. The dashed curve is based on the linear parameterization
of Randrup and Ko \cite{RAN80}, while the dotted curve is based on
the quartic parameterization of Zwermann \cite{ZWER87}. The experimental
data are from Ref. \cite{DATA}.

\item{Fig. 6:} Same as Fig. 4, for $pp\rightarrow p\Sigma ^0K^+$.

\item{Fig. 7:} Same as Fig. 4, for $pp\rightarrow p\Sigma ^+K^0$.

\item{Fig. 8:} Same as Fig. 4, for $pp\rightarrow n\Sigma ^+K^+$.

\item{Fig. 9:} Isospin-averaged cross section $\sigma _{NN\rightarrow
NYK}(=\sigma _{NN\rightarrow N\Lambda K}+
\sigma _{NN\rightarrow N\Sigma K}$). The solid curve is our model prediction
and the dashed curve is from the parameterization of Ref. \cite{RAN80}.

\item{Fig. 10:} Same as Fig. 9 for $\sigma _{N\Delta\rightarrow
NYK}(=\sigma _{N\Delta\rightarrow N\Lambda K}+
\sigma _{N\Delta\rightarrow N\Sigma K}$).

\item{Fig. 11:} Same as Fig. 9 for $\sigma _{\Delta\Delta\rightarrow
NYK}(=\sigma _{\Delta\Delta\rightarrow N\Lambda K}+
\sigma _{\Delta\Delta\rightarrow N\Sigma K}$).

\item{Fig. 12:} The energy distribution of baryon-baryon
collisions that are above the kaon production threshold.

\item{Fig. 13:} Kaon production probabilities from
different channels obtained in three calculations.

\item{Fig. 14:} Comparisons of kaon momentum spectra obtained in this work
using the theoretical kaon production cross sections (solid curve) with
the experimental data (open squares) from Ref. \cite{EXP3}.

\end{description}


\bigskip\bigskip

\begin{thebibliography}{99}

\bibitem{EXP1}S. Nagamiya, M. C. Lemair, E. Moeller, S. Snetzer, G.
Shapiro, H. Steiner, and I. Tanihata, Phys. Rev. C24 (1981) 971;\\
S. Snetzer, R. M. Lombard, M. C. Lemair, E. Moeller, S. Nagamiya, G.
Shapiro, H. Steiner, and I. Tanihata, {\it ibid.}
C40 (1989) 640.

\bibitem{EXP2} J. B. Carroll, Nucl. Phys. A488 (1988) 203c;\\
A. Shor, E. F. Barasch, J. B. Carroll, T. Hallman, C. Igo, G. Kalnins,
P. Kirk, G. F. Krebs, P. Lindstrom, M. A. McMahan, V. Perez-Mendez,
S. Trentalange, F. J. Urban, and Z. F. Wang, Phys. Rev. Lett.
63 (1989) 2192.

\bibitem{EXP3}E. Grosse, Prog. Part. Nucl.
Phys. 30 (1993) 89;\\
P. Senger {\it et al.,} Nucl. Phys. A573 (1993) 757c;\\
D. Miskowiec, {\it et al.,} Phys. Rev. Lett. 72
(1994) 3650.

\bibitem{EXP4} T. Abbot {\it et al.,} E802 Collaboration, Phys. Rev. Lett.
64 (1990) 847;\\
T. Abbot {\it et al.,} E802 Collaboration,
Phys. Rev. C50 (1994) 1024.

\bibitem{RAN80} J. Randrup and C. M. Ko, Nucl. Phys. A343 (1980) 519;
A411 (1983) 537.

\bibitem{CUG84} J. Cugnon and R. M. Lombard, Nucl. Phys. A422 (1984) 635.

\bibitem{AICH85} J. Aichelin and C. M. Ko, Phys. Rev.
Lett. 55 (1985) 2661.

\bibitem{AICH87} J. Aichelin, A. Rosenhauer, G. Peilert, H. St\"ocker,
and W. Greiner, Phys. Rev. Lett. 58 (1987) 1926.

\bibitem{ZWER87} B. Sch\"urmann and W. Zwermann, Phys. Lett. B183
(1987) 31;\\
W. Zwermann, Mod. Phys. Lett. A3 (1988) 251.

\bibitem{MOS91}W. Cassing, V. Metag, U. Mosel, and K. Niita, Phys. Rep.
188 (1990) 363;\\
A. Lang, W. Cassing, U. Mosel, and K. Weber, Nucl. Phys. A541
(1992) 507.

\bibitem{LI92} G. Q. Li, S. W. Huang, T. Maruyama, Y. Lotfy,
D. T. Khoa, and A. Faessler, Nucl. Phys.  A537
(1992) 645;\\
G. Q. Li, A. Faessler, and S. W. Huang, Prog. Part. Nucl. Phys.
31 (1993) 159.

\bibitem{HUANG93} S. W. Huang, A. Faessler, G. Q. Li, R. K. Puri,
E. Lehmann, M. A. Martin, and D. T. Khoa, Phys. Lett.
B298 (1993) 41.

\bibitem{FANG94} X. S. Fang, C. M. Ko, and Y. M. Zheng,
Nucl. Phys. A556 (1993) 499;\\
X. S. Fang, C. M. Ko, G. Q. Li, and Y. M. Zheng,
Phys. Rev. C49 (1994) R608;
Nucl. Phys. A575 (1994) 766.

\bibitem{MOS94} T Maruyama, W. Cassing, U. Mosel, S. Teis, and
K. Weber, Nucl. Phys. A573 (1994) 653.

\bibitem{LIBA94} B. A. Li, Phys. Rev. C50 (1994) 2144.

\bibitem{AICH94} G. Hartnack, J. Jaenicke, L. Sehn,
H. St\"ocker, and J. Aichelin, Nucl. Phys. A580 (1994) 643.

\bibitem{LI95} G. Q. Li and C. M. Ko, Phys. Lett. B, in press.

\bibitem{ko91} C. M. Ko, Z. G. Wu, L. H. Xia, and G. E. Brown, Phys.
Rev. Lett. 20, (1991) 2577; Phys. Rev. C43, 1881 (1991).

\bibitem{ARC} Y. Pang, T. J. Schlagel, and S. H. Kahana, Phys. Rev. Lett.
68 (1992) 2743.

\bibitem{RQMD} R. Mattiello, H. Sorge, H. St\"ocker, and W. Greiner,
Phys. Rev. Lett. 63 (1989) 1459.

\bibitem{gqli95}G. Q. Li, C. M. Ko, and B. A. Li, Phys. Rev. Lett.
74 (1995) 235; Phys. Rev. C., to be submitted.

\bibitem{kapl86} D. B. Kaplan and A. E. Nelson, Phys. Lett. B175 (1986) 57.

\bibitem{brow94} G. E. Brown, C. H. Lee, M. Rho, and V. Thorsson, Nucl.
Phys. A567 (1994) 937.

\bibitem{DATA} O. Benary, L. R. Price, and G. Alexander, Report No. UCRL-2000
NN, 1970;\\
A. Baldini, V. Flminio, W. G. Moorhead, and D. R. O. Morrison,
Total Cross-Sections for Reactions of High Energy Particles, (Springer-Verlag,
Berlin, 1988).

\bibitem{FAE94} K. Tsushima, S. W. Huang, and A. Faessler, Phys. Lett.
B337 (1994) 245;  ~J. Phys. G21 (1995) 33.

\bibitem{KO89} J. Q. Wu and C. M. Ko, Nucl. Phys. A499 (1989) 810.

\bibitem{LAGET91} J. M. Laget, Phys. Lett. B259 (1991) 24.

\bibitem{YAO} T. Yao, Phys. Rev. 125 (1961) 1048.

\bibitem{MARTIN76} B. R. Martin, Nucl. Phys. B94 (1975) 413.

\bibitem{XIONG90} L. Xiong, C. M. Ko, and J. Q. Wu, Phys. Rev. C42
(1990) 2231.

\bibitem{MOS95} W. Peters, U. Mosel, and A. Engel, Nucl. Phys. A,
in press.

\bibitem{brow75} G. E. Brown and W. Weise, Phys. Rep. 22 (1975) 279.

\bibitem{PAN81} B. Friedman, V. R. Pandharipande, and Q. N. Usmani,
Nucl. Phys. A372 (1981) 483.

\bibitem{BROWN88} G. F. Bertsch, G. E. Brown, V. Koch, and B. A. Li,
Nucl. Phys. A490 (1988) 745.

\bibitem{XIA89} C. M. Ko,  L. H. Xia, and P. J. Siemens, Phys. Lett.
B231 (1989) 16.

\bibitem{henn94}P. A. Henning and H. Umezawa, Nucl. Phys. A571
(1994) 617.

\bibitem{MACH87} R. Machleidt, K. Holinde, Ch. Elster, Phys. Rep.
149 (1987) 1;\\
R. Machleidt, Adv. Nucl. Phys. 19 (1989) 189.

\bibitem{WEISE} T. Ericson and W. Weise, {\it Pions and Nuclei},
(Oxford University Press, 1988) p. 31.

\bibitem{SWART} P. M. M. Maessen, Th. A. Rijken, and J. J. de Swart,
Phys. Rev. C40 (1989) 2226.

\bibitem{SPETH} A. Reuber, K. Holinde and J. Speth, Nucl. Phys. A570
(1994) 543.

\bibitem{KO87} C. M. Ko, Q. Li, and R. Wang, Phys. Rev. Lett. 59
(1987) 1084;\\
C. M. Ko and Q. Li, Phys. Rev. C37 (1988) 2270;\\
Q. Li, J. Q. Wu, and C. M. Ko, Phys. Rev. C39 (1989) 849;\\
C. M. Ko, Nucl. Phys. A495 (1989) 321c.

\bibitem{LI94B} G. Q. Li, C. M. Ko, X. S. Fang, and Y. M. Zheng, Phys.
Rev. C49 (1994) 1139;\\
G. Q. Li and C. M. Ko, Phys. Rev. C50 (1994) 1725.

\bibitem{LI94C} G. Q. Li, C. M. Ko, and X. S. Fang, Phys. Lett. B329
(1994) 149.

\end{thebibliography}
\end{document}